\documentclass[fleqn,twoside]{article}
\usepackage{espcrc2}

\usepackage{epsfig}
\usepackage{rotating}
\usepackage{ifthen,array}
 \usepackage{setspace}
 \usepackage{fancyhdr}
 \usepackage{moreverb}
 \usepackage{rotating}
\usepackage{graphicx}
\def\e6{$E(6)$}
\def\10{$SO(10)$}
\def\21{$SU(2) \otimes U(1) $}

\def\422{$SU(4) \otimes SU(2) \otimes SU(2)$}
\def\321{$SU(3) \otimes SU(2) \otimes U(1)$}
\def\lsim{\raise0.3ex\hbox{$\;<$\kern-0.75em\raise-1.1ex\hbox{$\sim\;$}}}
\def\gsim{\raise0.3ex\hbox{$\;>$\kern-0.75em\raise-1.1ex\hbox{$\sim\;$}}}
\def\lfv{lepton flavour violation }
\def\lnv{lepton number violation }
\def\meff{\langle m_{\nu} \rangle}
\newcommand{\ed}{\end{document}}
\DeclareMathAlphabet{\mathsc}{OT1}{cmr}{m}{sc}

\newcommand{\CL}   {C.L.}
\newcommand{\dof}  {d.o.f.}

\newcommand{\eVq}  {\rm{eV}^2}
\newcommand{\Sol}  {\mathsc{sol}}
\newcommand{\Atm}  {\mathsc{atm}}

\newcommand{\Dms}  {\Delta m^2_\Sol}
\newcommand{\Dma}  {\Delta m^2_\Atm}

\def \znbb {$0\nu\beta\beta$ }

\def\meff{\langle m_{\nu} \rangle}

\let\vev\VEV

\def\e6{$E(6)$}
\def\10{$SO(10)$}
\def\21{$SU(2) \otimes U(1) $}

\def\422{$SU(4) \otimes SU(2) \otimes SU(2)$ }
\def\321{$SU(3) \otimes SU(2) \otimes U(1)$ }

\renewcommand{\baselinestretch}{1.068}

\newcommand{\AmS}{{\protect\the\textfont2
  A\kern-.1667em\lower.5ex\hbox{M}\kern-.125emS}}

\title{Concluding talk at NOW 2006}

\author{J. W. F. Valle\address{AHEP Group, Instituto de F\'isica
    Corpuscular -- C.S.I.C./Universitat
    de Val\`encia\\
    \small\it Edificio Institutos de Paterna, Apt 22085, E-46071,
    Val\`encia, Spain}%
  \thanks{Work supported by MEC grants FPA2005-01269 and
    FPA2005-25348-E, by Generalitat ACOMP06/154, and by the EC RTN
    Contract MRTN-CT-2004-503369 and ILIAS/N6 Contract
    RII3-CT-2004-506222. Thanks to M. Tortola, M. Maltoni and T. Schwetz.}  }
\begin{document}

\begin{abstract}
  This writeup summarizes the status of neutrino oscillations,
  including recent fluxes and experimental data, as of summer 2006.
  A discussion is given on the current status of absolute scale of
  neutrino mass from tritium, \znbb and cosmological observations, as
  well as the prospects for the next generation of experiments,
  including \lfv searches, and their theoretical significance.
\vspace{.3pc}
\end{abstract}
%% keywords      {neutrino oscillations, neutrino mass, astroparticle physics}
\maketitle

\section{Intoduction}

With almost 30 plenary talks and 70 parallel session talks the task of
summarizing all of them all into a single talk is an impossible
mission. I will try instead simply to highlight some aspects of the
talks that touched my own prejudices.

The progress in the physics of neutrino oscillations in the last few
years has been truly remarkable.
Oscillations are now established, implying that neutrinos have masses,
as first suggested by theorists in the early eighties, both on general
grounds~\cite{Weinberg:1980bf,schechter:1980gr} and on the basis of
various versions of the seesaw mechanism~\cite{Valle:2006vb}.
This is a profound discovery that marks the beginning of a new age in
neutrino physics. 

A gold rush towards precision results has been initiated, whose aim is
to probe $\theta_{13}$, to study leptonic CP violation and determine
the nature of neutrinos. Hopefully this will shed light on the
ultimate origin the universe and certainly that of neutrino mass.

\section{Data}

Thanks to the accumulation of events over a wide range of energy, and
to the measurement of the dip in the L/E (neutrino flight length L
over neutrino energy E) distribution of the muon neutrino
disappearance probability, the interpretation of the atmospheric data
has finally turned into an unambiguous signal of $\nu_\mu
\leftrightarrow \nu_\tau$ oscillations, marking the beginning of a new
era.

The interpretation of solar data {\sl per se} is still ambiguous, with
viable alternative explanations involving spin flavour
precession~\cite{barranco:2002te,miranda:2000bi} or non-standard
neutrino interactions~\cite{guzzo:2001mi}.
Results on (or relevant to) solar neutrinos were presented here by
Broggini, Maneira, Pulido, Raghavan, Ranucci, Serenelli and Smy.
Reactor neutrino data from KamLAND not only confirm the solar neutrino
deficit but also observe the spectrum distortion as expected for
oscillations.  Reactors have played an important role in establishing
the robustness of the neutrino flavor oscillation interpretation {\sl
  vis a vis} the existence of solar density
fluctuations~\cite{Burgess:2003su} in the solar radiative zone as
produced by random magnetic fields~\cite{Burgess:2003fj}, and also
with respect to the effect of convective zone magnetic fields, should
neutrinos posses nonzero neutrino transition magnetic
moments~\cite{Miranda:2003yh}.  Within the oscillation picture KamLAND
has also identified large mixing angle oscillation as its ``unique''
solution, ``solving'', in a sense, the solar neutrino problem. Note
however that the interpretation of solar data {\sl vis a vis} neutrino
non-standard interactions~\cite{Miranda:2004nb} is not yet so robust.

A lot more is to come from the upcoming reactor experiments starting
with Double-Chooz, and a new series of proposed experiments such as
Daya-bay, RENO, Kaska, Angra, nicely reviewed in the talk by Cabrera.

Accelerators K2K \& MINOS confirm the atmospheric neutrino deficit as
well as a distortion of the energy spectrum consistent with the
oscillation hypothesis. More is to come from MINOS and the upcoming
experiments CNGS/OPERA, T2K, NOVA, as reported here by Gugliemi,
Kajita, Kato, Kopp, Rebel, Sioli, and others.

\section{Oscillation parameters}

The basic tool to interpret neutrino data is the lepton mixing matrix,
whose simplest unitary 3-dimensional form is given as a product of
effectively complex $2\times 2$ matrices~\cite{schechter:1980gr}
\begin{equation}
  \label{eq:2227}
K =  \omega_{23} \omega_{13} \omega_{12}
\end{equation}
where each factor is given as
\begin{equation}
   \label{eq:w13}
\omega_{13} = \left(\begin{array}{ccccc}
c_{13} & 0 & e^{i \Phi_{13}} s_{13} \\
0 & 1 & 0 \\
-e^{-i \Phi_{13}} s_{13} & 0 & c_{13}
\end{array}\right)\,,
 \end{equation}
 in the most convenient ordering chosen in the PDG~\cite{Yao:2006px}
 (here $c_{ij} \equiv \cos\theta_{ij}$ and $s_{ij} \equiv
 \sin\theta_{ij}$).
 The two Majorana phases associated to 12 and 23 can be eliminated
 insofar as neutrino oscillations are concerned, since they only
 affect lepton-number violating processes, like \znbb. Thus one can
 take the 12 and 23 factors as real, $ \omega_{23} \to r_{23}$ and $
 \omega_{12} \to r_{12}$. There is then a unique CP phase $\Phi_{13}$,
 analogous to the KM phase $\delta$ of quarks, that will be studied in
 future oscillation experiments, such as T2K and NOVA.
 Currently, however, oscillations have no sensitivity to this phase
 and we will neglect it in the analysis of current data.

 In such approximation oscillations depend on the three mixing
 parameters $\sin^2\theta_{12}, \sin^2\theta_{23}, \sin^2\theta_{13}$
 and on the two mass-squared splittings $\Dms \equiv \Delta m^2_{21}
 \equiv m^2_2 - m^2_1$ and $\Dma \equiv \Delta m^2_{31} \equiv m^2_3 -
 m^2_1$ characterizing solar and atmospheric neutrinos.  The fact that
 $\Dms \ll \Dma$ implies that one can set $\Dms \to 0$ in the analysis
 of atmospheric and accelerator data, and $\Dma$ to infinity in the
 analysis of solar and reactor data.

\subsection{Present status}
 \label{sec:present-status}

\begin{figure}[t] \centering
    \includegraphics[width=\linewidth,height=7cm]{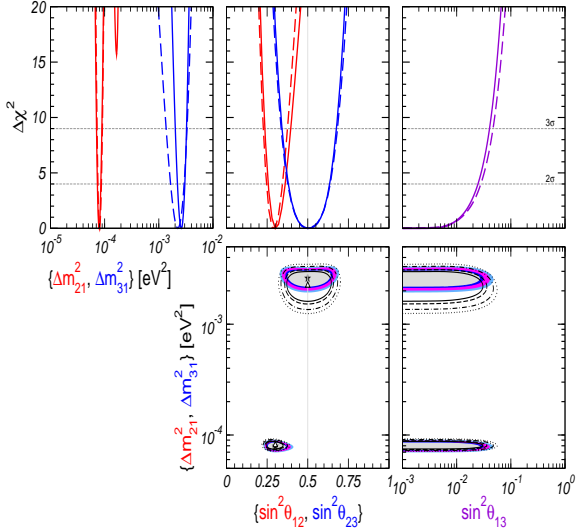}
    \caption{\label{fig:global} %
      Current 90\%, 95\%, 99\%, and 3$\sigma$ \CL\ neutrino
      oscillation regions for 2 \dof\, as of summer 2006,
      from~\cite{Maltoni:2004ei}. In top panels $\Delta \chi^2$ is
      minimized with respect to undisplayed parameters.}
\end{figure}

 The current three--neutrino oscillation parameters are summarized in
 Fig.~\ref{fig:global}. Equivalent results by the Bari group are in
 excellent agreement with those reported here, both
 pre~\cite{Fogli:2005cq} and post-MINOS~\cite{Fogli:2006yq}.  The
 analysis employs the latest Standard Solar
 Model~\cite{Bahcall:2005va} which we heard here in the talk by
 Serenelli, and includes all new neutrino oscillation data from SNO
 salt~\cite{Aharmim:2005gt}, K2K~\cite{Ahn:2006zz} and
 MINOS~\cite{Michael:2006rx}, described in Appendix C of
 hep-ph/0405172 (v5)~\cite{Maltoni:2004ei}~\footnote{ In addition to
   good calculations of the neutrino
   fluxes~\cite{Bahcall:2004fg,Honda:2004yz}, cross sections and
   response functions, we need an accurate description of neutrino
   propagation in the Sun and the Earth, including matter
   effects~\cite{mikheev:1985gs,wolfenstein:1978ue}.  }.

 The upper panels of the figure show $\Delta \chi^2$ as a function of
 the three mixing parameters $\sin^2\theta_{12}, \sin^2\theta_{23}, \,
 \sin^2\theta_{13}$ and two mass squared splittings $\Delta m^2_{21},
 \Delta m^2_{31}$, minimized with respect to the undisplayed
 parameters. The lower panels give two-dimensional projections of the
 allowed regions in five-dimensional parameter space.  In addition to
 a confirmation of oscillations with $\Dma$, accelerator neutrinos
 provide a better determination of $\Dma$ as one can see by comparing
 dashed and solid lines in Fig.~\ref{fig:global}.  The recent MINOS
 data~\cite{Michael:2006rx} lead to an improved determination and a
 slight increase in $\Dma$.
As already mentioned, reactors~\cite{araki:2004mb} have played a
crucial role in selecting large-mixing-angle (LMA)
oscillations~\cite{pakvasa:2003zv} out of the previous ``zoo'' of
possible solar neutrino oscillation
solutions~\cite{Gonzalez-garcia:2000sq,fogli:2001vr}.
Table~\ref{tab:summary} gives the current best fit values and allowed
3$\sigma$ ranges of oscillation parameters.
\begin{table}[t] \centering    \catcode`?=\active \def?{\hphantom{0}}
      \begin{tabular}{|l|c|c|}        \hline        parameter & best
      fit & 3$\sigma$ range         \\  \hline\hline        $\Delta
      m^2_{21}\: [10^{-5}~\eVq]$        & 7.9?? & 7.1--8.9 \\
      $\Delta m^2_{31}\: [10^{-3}~\eVq]$ & 2.6?? &  2.0--3.2 \\
      $\sin^2\theta_{12}$        & 0.30? & 0.24--0.40 \\
      $\sin^2\theta_{23}$        & 0.50? & 0.34--0.68 \\
      $\sin^2\theta_{13}$        & 0.00 & $\leq$ 0.040 \\
      \hline
\end{tabular}    \vspace{2mm}
\caption{\label{tab:summary} Neutrino oscillation parameters as of Summer 2006, 
from Ref.~\cite{Maltoni:2004ei}.}
\end{table}

Note that CP violation disappears in a three--neutrino scheme when two
neutrinos become degenerate or when one of the angles
vanishes~\cite{schechter:1980bn}.  As a result CP violation is doubly
suppressed, first by $\alpha \equiv \Dms/\Dma$ and also by the small
value of $\theta_{13}$.
\begin{figure}[t]
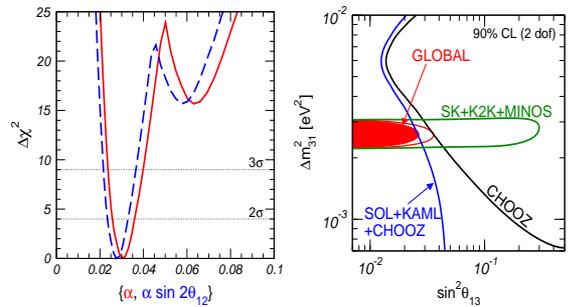
 \centering
  \includegraphics[height=4cm,width=.48\linewidth]{F-fcn.alpha06.eps}
\includegraphics[height=4cm,width=.48\linewidth]{th13-06.eps}
\caption{\label{fig:alpha}%
  $\alpha \equiv \Dms / \Dma$ and $\sin^2\theta_{13}$ bound from the
  updated analysis given in Ref.~\cite{Maltoni:2004ei}.}
\end{figure}
The left panel in Fig.~\ref{fig:alpha} gives the parameter $\alpha$,
while the right panel shows the impact of different data samples on
constraining $\theta_{13}$.  One sees that for larger $\Dma$ values
the bound on $\sin^2\theta_{13}$ is dominated by CHOOZ, while for low
$\Dma$ the solar and KamLAND data become quite relevant.

\subsection{Robustness}
\label{sec:robustness}

Reactor neutrino data have played a crucial role in testing the
robustness of solar oscillations vis a vis astrophysical
uncertainties, such as magnetic fields in the
radiative~\cite{Burgess:2003su,Burgess:2003fj} and convective
zone~\cite{miranda:2000bi,guzzo:2001mi,barranco:2002te}, leading to
stringent limits on neutrino magnetic transition
moments~\cite{Miranda:2003yh}.
KamLAND has also played a key role in identifying oscillations as
``the'' solution to the solar neutrino problem~\cite{pakvasa:2003zv}
and also in pinning down the LMA parameter region among previous wide
range of oscillation
solutions~\cite{Gonzalez-garcia:2000sq,fogli:2001vr}.

\begin{figure}[t] \centering
    \includegraphics[height=3cm,width=.55\linewidth]{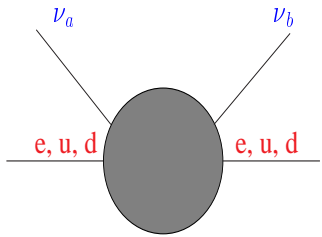}
    \caption{\label{fig:nuNSI} %
      Non-standard neutrino interactions arise, e.~g., from the
      non-unitary structure of charged current weak interactions
      characterizing seesaw-type schemes~\cite{schechter:1980gr}.}
\end{figure}

However, there is still some fragility in the interpretation of the
data in the presence of sub-weak strength ($\sim \varepsilon G_F$)
non-standard neutrino interactions (NSI) (Fig.~\ref{fig:nuNSI}).
Indeed, most neutrino mass generation mechanisms imply the existence
of such dimension-6 operators. They can be of two types:
flavour-changing (FC) and non-universal (NU). Their presence leads to
the possibility of resonant neutrino conversions even in the absence
of neutrino masses~\cite{valle:1987gv}.
While model-dependent, the expected NSI magnitudes may well
fall within the range that will be tested in future precision
studies~\cite{Huber:2004ug}.
For example, in the inverse seesaw model~\cite{Deppisch:2004fa} the
non-unitary piece of the lepton mixing matrix can be sizeable, hence
the induced non-standard interactions.  Relatively sizeable NSI
strengths may also be induced in supersymmetric unified
models~\cite{hall:1986dx} and models with radiatively induced neutrino
masses~\cite{zee:1980ai,babu:1988ki}.

The determination of atmospheric neutrino parameters $\Dma$ and
$\sin^2\theta_\Atm$ is hardly affected by the presence of NSI on
down-type quarks~\cite{fornengo:2001pm}. 

In contrast, the determination of solar neutrino parameters is not
quite robust against the existence of NSI~\cite{Miranda:2004nb}, even
if reactor data are included. 
The issue can only be resolved by future low and intermediate energy
solar neutrino data mentioned by Raghavan, with enough precision to
sort out the detailed profile of the solar neutrino conversion
probability.
 
\subsection{Future prospects}
\label{sec:future-prospects}

\begin{figure}[t] \centering
\includegraphics[height=6cm,width=.7\linewidth]{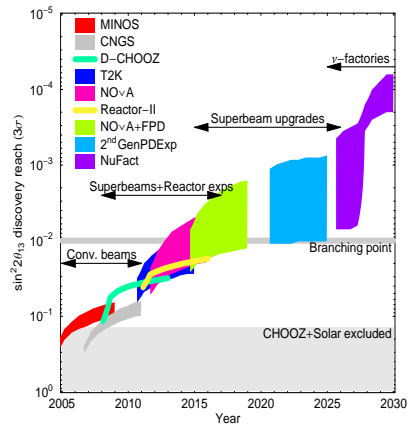}
\caption{\label{fig:t13fut} The hunt for $\theta_{13}$: artist's view
  of anticipated sensitivities on $\theta_{13}$ given in
  Ref.~\cite{Albrow:2005kw}.}
\end{figure}

Upcoming reactor and accelerator long baseline experiments aim at
improving the sensitivity on $\sin^2\theta_{13}$~\cite{Huber:2004ug}.
The value of $\theta_{13}$ is a key input and the start of an
ambitious long-term effort towards probing CP violation in neutrino
oscillations~\cite{Alsharoa:2002wu,apollonio:2002en,albright:2000xi}.
Prospects of accelerator and reactor neutrino oscillation experiments
for the coming years have been extensively discussed in the literature
and there have been several talks at this conference, for example
those of Declais, Huber, Kajita, Kato, Kopp, Lindner, Nunokawa and
Schwetz. Here I simply illustrate in Fig.~\ref{fig:t13fut} the
anticipated evolution of the $\theta_{13}$ discovery reach for the
global neutrino program, see details in Ref.~\cite{Albrow:2005kw}.
                                    
One important comment is that even a small residual non-standard
interaction in this ``solar'' (e-tau) channel has dramatic
consequences for the sensitivity loss for $\theta_{13}$ at a neutrino
factory~\cite{huber:2001de}.
To make these experiments meaningful it is a must to have a near
detector capable of sorting out for NSI with high sensitivity.

In contrast, future neutrino factories will probe flavor changing
non-standard neutrino-matter interactions in the ``atmospheric''
(mu-tau) channel with sensitities which are substantially improved
with respect to current ones~\cite{huber:2001zw}.  For example, a 100
GeV NUFACT can probe these at the level of $|\epsilon| < \mathrm{few}
\times 10^{-4}$ at 99 \% C.L.
 
Improving the sensitivities on NSI constitutes at a near detector is a
necessary pre-requisit. In short, probing for NSI is an important item
and a window of opportunity for neutrino physics in the precision age.

To close this section let me mention that day/night effect studies in
large water Cerenkov solar neutrino experiments such as UNO, Hyper-K
or LENA has also been suggested as an alternative way to probe
$\theta_{13}$~\cite{Akhmedov:2004rq}.

\section{Lepton flavour violation}
\label{sec:lept-flav-viol}

\begin{figure}[h]
\centering
\includegraphics[clip,height=4.5cm,width=0.8\linewidth]{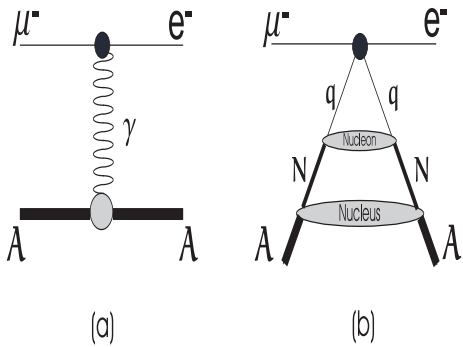}
\caption{Contributions to the nuclear $\mu^-- e^-$ conversion: (a)
  long-distance and (b) short-distance. For numerical results see
  Ref.~\cite{Deppisch:2004fa,Deppisch:2005zm}}
     \label{fig:Diagrams}
\end{figure}

\begin{figure}[h] \centering
\includegraphics[height=4cm,width=.8\linewidth]{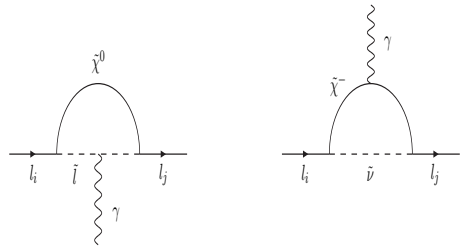}
\includegraphics[width=\linewidth,height=5cm]{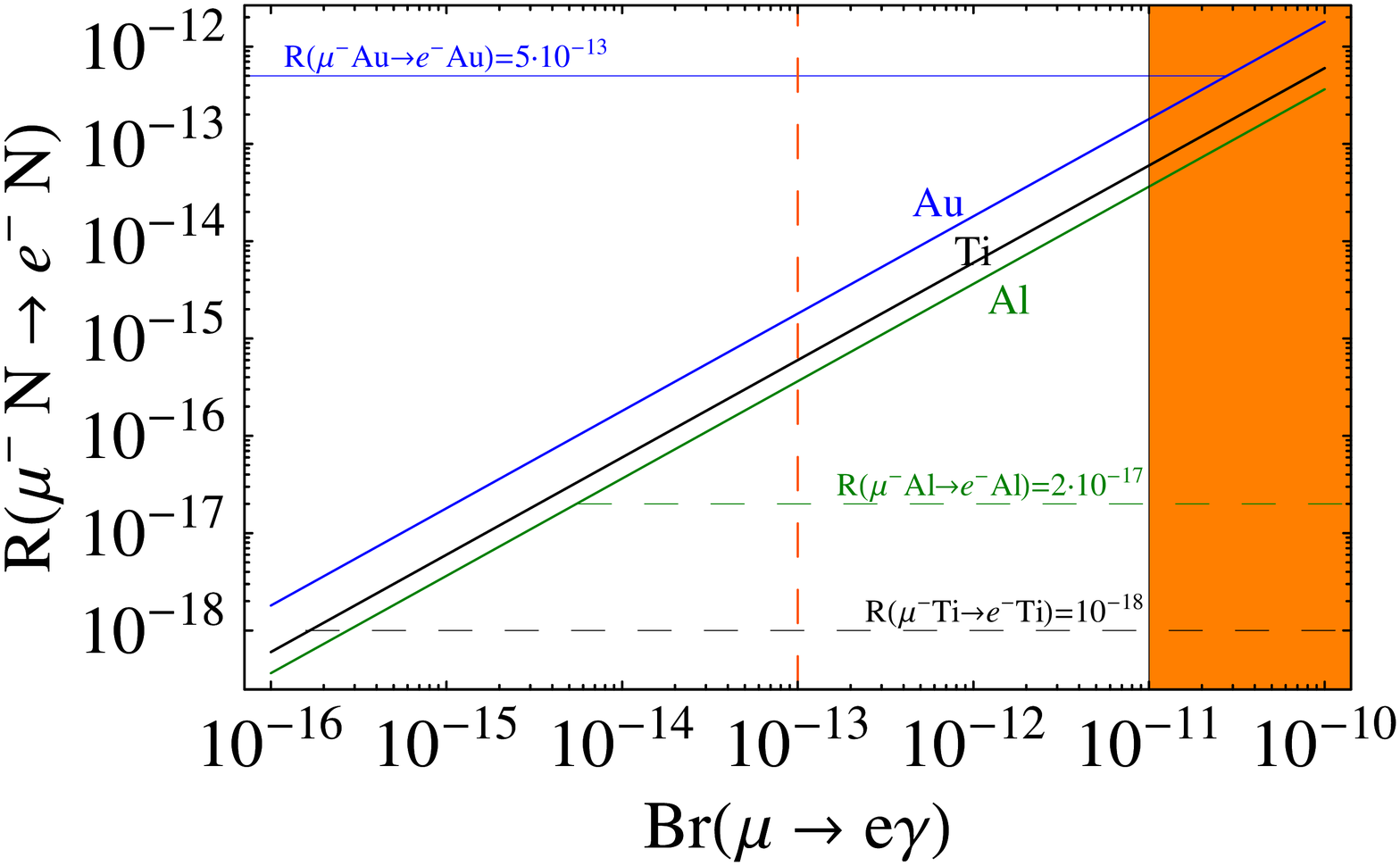} 
\vglue -1cm
\caption{\label{fig:mueg} Supersymmetric Feynman diagrams for \lfv
  and correlation between mu-e conversion and \(Br(\mu\to e\gamma)\)
  in the supersymmetric inverse seesaw model of
  Ref.~\cite{Deppisch:2004fa}.}
\end{figure}

The discovery of neutrino oscillations demonstrates that lepton
flavour conservation is not a fundamental symmetry of nature. It is
therefore natural to expect that it may show up elsewhere, for example
\(\mu\to e\gamma\) or nuclear $\mu^-- e^-$ conversion, as seen in
Fig.~\ref{fig:Diagrams}.
Indeed, in seesaw-type schemes of neutrino mass, \lfv is induced
either from neutral heavy lepton
exchange~\cite{Bernabeu:1987gr,Ilakovac:1994kj} as discussed here by
Vogel, or through the exchange of charginos (neutralinos) and
sneutrinos (charged sleptons) as discussed here by Masiero
~\cite{Hall:1985dx,borzumati:1986qx,casas:2001sr,Antusch:2006vw} (see
top panel in Fig.~\ref{fig:mueg}).
As illustrated in Fig.~\ref{fig:mueg} the rates for both processes can
be sizeable and fall within the sensitivity of upcoming experiments.
The calculation in Fig.~\ref{fig:mueg} is performed in the framework
of the generalized supersymmetric seesaw model of
Ref.~\cite{Deppisch:2004fa} to where I address you if you wish to
understand the interplay of neutral heavy
lepton~\cite{Bernabeu:1987gr} and supersymmetric contributions.
If the neutral heavy leptons are in the TeV range (a situation not
realizable in the minimal seesaw mechanism), the \(Br(\mu\to
e\gamma)\) rate can be enhanced even in the {\sl absence} of
supersymmetry. In this case the neutral heavy leptons that mediate
\lfv may be directly produced at accelerators~\cite{Dittmar:1990yg}.

Fig.~\ref{fig:mueg} also illustrates the correlation between nuclear
$\mu^--e^-$ conversion and $\mu^-\to e^-\gamma$ decay in the inverse
seesaw model for the nuclei Au, Ti, Al.  The shaded area and vertical
line denote the current bound and future sensitivity (PSI) on
$Br(\mu\to e\gamma)$, respectively.  The horizontal lines denote the
current bound (Au/SINDRUM II) and expected future sensitivities
(Al/MECO, Ti/PRISM) on $R(\mu^-N\to eN)$. More details in
Ref.~\cite{Deppisch:2005zm}.

Note also that since \lfv and CP violation can occur in the massless
neutrino limit, hence the allowed rates need not be suppressed by the
smallness of neutrino
masses~\cite{Bernabeu:1987gr,Ilakovac:1994kj,branco:1989bn,rius:1990gk}.

\section{Absolute scale of neutrino mass}
\label{sec:neutr-double-beta}

\begin{figure}[h]
  \centering
\includegraphics[clip,width=.46\linewidth,height=3.5cm]{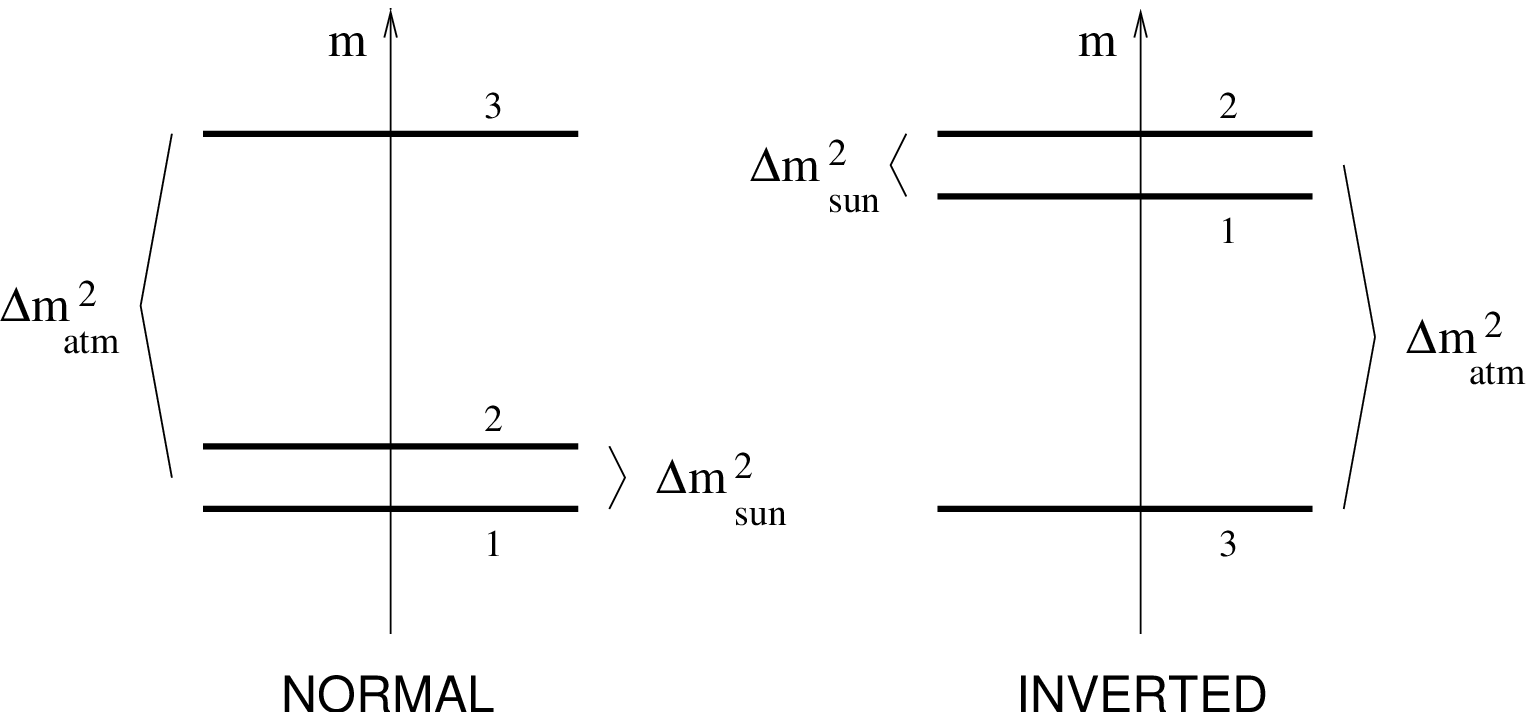}
\includegraphics[height=3.5cm,width=.51\linewidth]{plot-mi-m3.eps}
\caption{Which spectrum?}
 \label{fig:Which}
\end{figure}

Neutrino oscillations are insensitive to absolute masses and can not
probe whether neutrinos are Dirac or Majorana. Current data can not
determine whether the spectrum is normal or inverted, as illustrated
in Fig.~\ref{fig:Which}.
To settle the issue one needs kinematical tests, such as beta decay
studies~\cite{Drexlin:2005zt}, as discussed here by Sisti and
Weinheimer. The upcoming high precision neutrino mass experiment
KATRIN scales up both the size \& source intensity, aiming at a
sensitivity one order of better than that of the current Mainz-Troitsk
experiments.

Neutrinoless double beta decay and other \lnv processes, such as
neutrino transition electromagnetic
moments~\cite{schechter:1981hw,Wolfenstein:1981rk}
\cite{pal:1982rm,kayser:1982br} can probe the basic nature of
neutrinos.

\begin{figure}[h]
  \centering
\includegraphics[width=6cm,height=3.2cm]{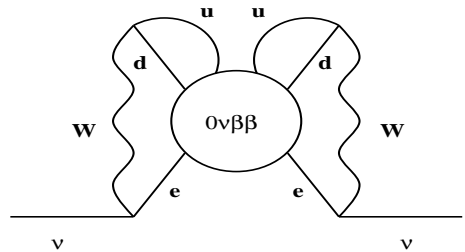}
\caption{Neutrinoless double beta decay and Majorana mass are
  equivalent~\cite{Schechter:1982bd}.}
 \label{fig:bbox}
\end{figure}

The significance of neutrinoless double beta decay stems from the
fact that, in a gauge theory, irrespective of the mechanism that
induces \znbb, it necessarily implies a Majorana neutrino
mass~\cite{Schechter:1982bd}, as illustrated in Fig.  \ref{fig:bbox}.
Thus it is a basic issue. Quantitative implications of the
``black-box'' argument are model-dependent, but the theorem itself
holds in any ``natural'' gauge theory.  

%%%% 

\begin{figure}[h]
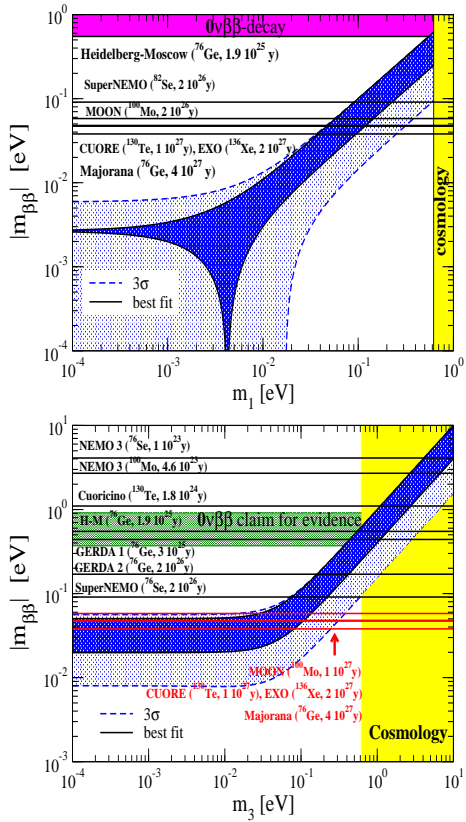

 \centering
\includegraphics[clip,width=.8\linewidth,height=5.4cm]{nh_paper.eps}
\includegraphics[clip,width=.8\linewidth,height=5.4cm]{ih_paper.eps}
 \caption{\znbb amplitude versus current oscillation data,
   from Ref.~\cite{Bilenky:2004wn}.}
\label{fig:nbbfut}
\end{figure}

Now that oscillations are experimentally confirmed we know that \znbb
must be induced by the exchange of light Majorana neutrinos, the
so-called "mass-mechanism". The \znbb amplitude depends on the 3
masses, 2 mixing angles, and 2 CP phases. Hence it involves the
absolute scale of neutrino mass, as well as the Majorana
phase~\cite{schechter:1980gr}, neither of which can be probed in
oscillations~\cite{bilenky:1980cx,Schechter:1981gk,doi:1981yb}.
The phenomenological situation was described here by Avignone,
Bettini, Fiorini, Pavan, Simkovik, Vala and Vogel. It clearly
distinguishes between normal and inverted hierarchical spectra, as
seen in Fig.~\ref{fig:nbbfut}: in the former hierarchy case there is
in general no lower bound on the \znbb rate, since there can be a
destructive interference amongst the neutrino amplitudes. In contrast,
the inverted neutrino mass hierarchy implies a ``lower'' bound for the
\znbb amplitude.
The best current limit on $\meff$ comes from the Heidelberg-Moscow
experiment. The current claim~\cite{Klapdor-Kleingrothaus:2004wj} (see
also Ref.~\cite{Aalseth:2002dt}) and the sensitivities of the upcoming
experiments are indicated in the compilation, courtesy of Simkovik,
displayed in Fig. \ref{fig:nbbfut}. It shows the estimated average
mass parameter characterizing the neutrino exchange contribution to
\znbb versus the lightest and heaviest neutrino masses.  The
calculation takes into account the current neutrino oscillation
parameters in \cite{Maltoni:2004ei} and state-of-the-art nuclear
matrix elements~\cite{Bilenky:2004wn}.
The upper (lower) panel corresponds to the cases of normal (inverted)
neutrino mass spectra. In these plots the ``diagonals'' correspond to
the case of quasi-degenerate
neutrinos~\cite{babu:2002dz}~\cite{caldwell:1993kn}~\cite{ioannisian:1994nx}.

\begin{figure}[h] \centering
    \includegraphics[width=.6\linewidth,height=3.5cm]{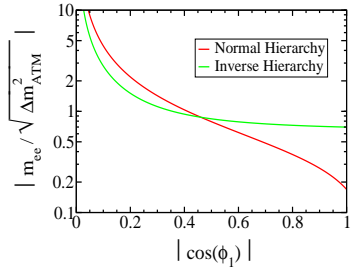}
    \caption{\label{fig:bbn-a4} %
      Lower bound on $|\vev{m_{ee}}|/\Dma$ vs $|\mathrm{cos}
      (\phi_1)|$ where $\phi_1$ is a Majorana phase. The lines in dark
      (red) and grey (green) correspond to normal and inverse
      hierarchy. Model of Ref.~\cite{Hirsch:2005mc}.}
\end{figure}
We now give two examples of model \znbb expectations. First,
Ref.~\cite{Hirsch:2005mc} proposes a specific normal hierarchy model
for which a lower bound on \znbb can be placed, as a function of the
value of the Majorana violating phase $\phi_1$, as indicated in
Fig.~\ref{fig:bbn-a4}.  Second, the $A_4$ model~\cite{babu:2002dz}
gives a lower bound on the absolute Majorana neutrino mass
$m_{\nu}\gsim 0.3$ eV and may therefore be tested in \znbb searches.

%%% ggg

\begin{figure}[h] \centering
\includegraphics[height=4cm,width=.45\textwidth]{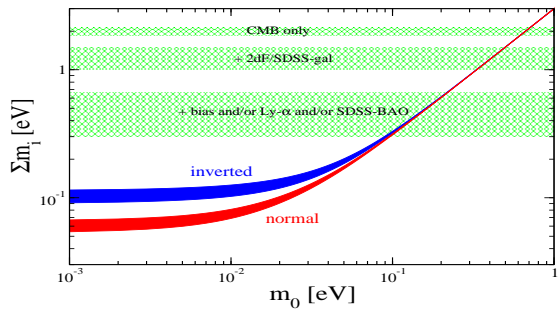}
\caption{\label{fig:sergio} %
  Sensitivity of cosmology to neutrino mass, see
  Ref.~\cite{Lesgourgues:2006nd}.}
\end{figure}

The absolute scale of neutrino masses will be tested by its effect on
the cosmic microwave background and the large scale structure of the
Universe, as discussed here by Elgaroy, Palazzo, Pastor and
Viel~\cite{Lesgourgues:2006nd,Hannestad:2006zg,Fogli:2006yq}, and
illustrated in Fig.~\ref{fig:sergio}.

%---------------------------------------------------------------------

\section{Neutrinos as probes }
\label{sec:neutrinos-as-probes}

Not only neutrino properties can be probed via cosmology and
astrophysics, but also, once well-determined, they can be used as
astro-probes (Sun, Supernovae, pulsars, etc), cosmo-probes even
geo-probes. Here there were many related talks by Elgaroy, Mangano,
Pastor, Villante, and Viel. Neutrinos affect nucleosynthesis, 
large scale structure, the CMB and possibly the generation of 
the matter anti-matter asymmetry.

Like photons, cosmic rays and gravitational waves, neutrinos are one
of the basic messengers of the Big Bang capable of probing early
stages of its evolution. In the leptogenesis scenarios (discussed here
by Akhmedov, di Bari, Ma and Petcov) neutrinos could probe the
Universe even down to epochs prior to the electroweak phase
transition.

Neutrinos are also basic probes is astrophysics. Having only weak
interactions, they are ideal to monitor the interior of stars, such as
the the Sun.

Supernova neutrinos were discussed by Cardall, Fleurot, Lunardini and
Vagins. The measurement of a large number of neutrinos from a future
galactic supernova neutrino signal will give us important information
on supernova parameters. Here I give a simplified plot (small
$\theta_{13}$ approximation) taken from Ref.~\cite{Minakata:2001cd} as
an illustration of the potential to probe astrophysics from a precise
knowledge of neutrino properties.
\begin{figure}[h] \centering
\includegraphics[height=4cm,width=.48\linewidth]{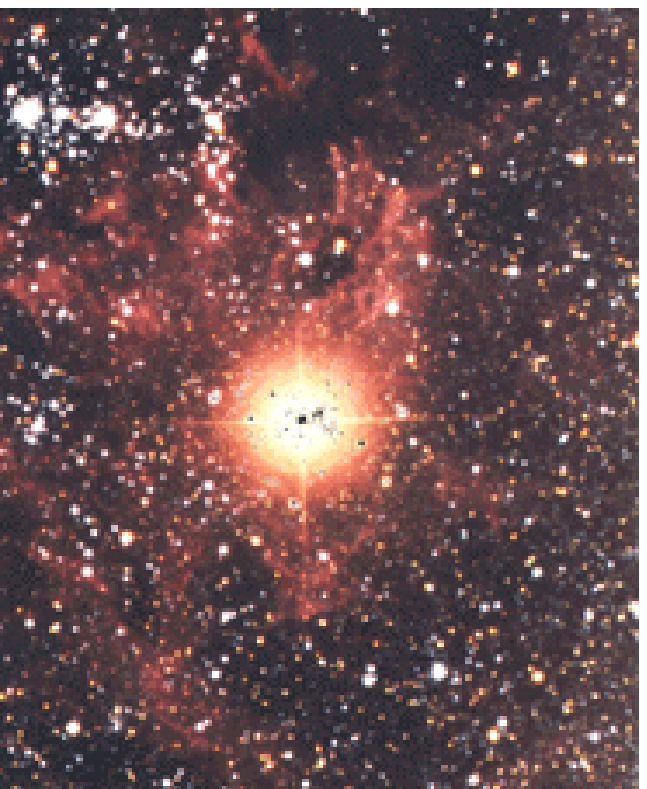}
\includegraphics[width=.49\linewidth,height=4.2cm]{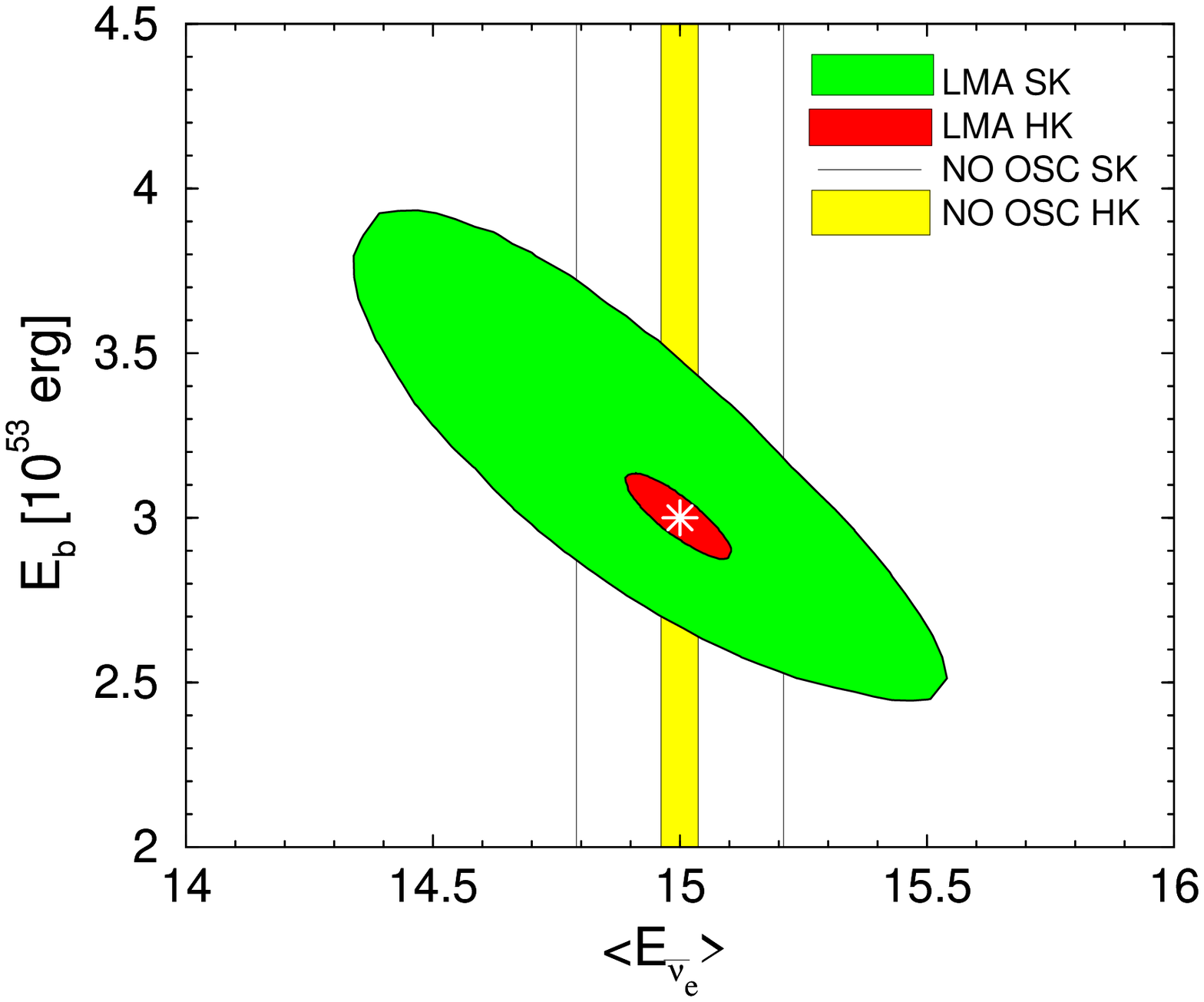} 
\vglue -.5cm     
\caption{\label{fig:sn} Improved supernova parameter determination
  attainable from the neutrino signal from 10 kpc galactic supernova
  within future neutrino telescopes Hyper-K versus Super-K.}
\end{figure}

Let me also mention that new effects in neutrino conversions at the
core of supernovae (neutron-rich regime) are expected when neutrinos
have non-standard
interactions~\cite{valle:1987gv,Nunokawa:1996tg,Nunokawa:1996ve}.
These could induce new inner resonant conversions, over and above
those that arise from oscillations.

Before closing this section, let me mention that neutrinos are ideal
probes of the high energy Universe, discussed here by Billoir, Karle,
Flaminio, Stanev and Sigl. For example, one expects high energy
neutrinos from AGNs and GRBs. Typically the accelerated primaries make
pions, leading to comparable fluxes of neutrinos and gammas due to
isospin. In contrast to gamma-rays, the neutrino spectrum is
essentially unmodified. One set of observables to monitor are the
flavor ratios, which are sensitive both to neutrino oscillations as
well as neutrino non-standard interactions.

Here I give a very useful roadmap-plot presented by Sigl (see his talk
at these proceedings for details). This plot makes extrapolations as
to what high energy neutrino fluxes could be on the basis of the
gamma-flux constraints (e.~g. by EGRET) at lower energies.
\begin{figure}[h] \centering
\includegraphics[height=6.5cm,width=\linewidth]{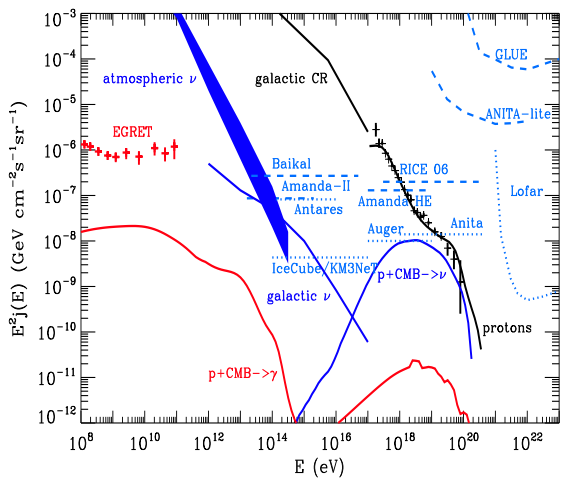}
\vglue -.8cm     
  \caption{\label{fig:road}
Sigl's map to high energy astrophysics.}
\end{figure}

%-----------------------------------------------------------------------

\section{Origin of neutrino mass}
\label{sec:origin-neutrino-mass}

This is one of the most well-kept secrets of nature.  Gauge theories
prefer Majorana neutrinos (see historical talk by
Esposito)~\cite{Weinberg:1980bf,schechter:1980gr} irrespective the
detailed mechanism of neutrino mass generation. While possible, the
emergence of Dirac neutrinos would constitute a surprise, indicating
the existence of an accidental lepton number symmetry whose
fundamental origin should be understood.  There are some ideas for
generating light Dirac neutrinos. For example, theories involving
large extra dimensions offer a novel scenario to account for small
Dirac neutrino masses. Within this picture, right-handed neutrinos
propagate in the bulk, while left-handed neutrinos, being a part of
the lepton doublet, live only on the Standard Model (SM) branes. In
this picture neutrinos get naturally small Dirac masses via mixing
with a bulk fermion.

However neutrinos  are more likely Majorana.

In contrast to SM charged fermions, neutrinos do not get masses after
the electroweak symmetry breaks through by the nonzero vacuum
expectation value (vev) of the Higgs scalar doublet, since they come
in just one chiral species. There is, however, an effective lepton
number violating dimension-five operator $LL\Phi\Phi$ (where L denotes
any of the lepton doublets and $\Phi$ the Higgs) which can be added to
the SM~\cite{Weinberg:1980bf}.  After the Higgs mechanism this
operator induces Majorana neutrino masses, thus providing a natural
way to account for their smallness, irrespective of their specific
origin. Little more can be said from first principles about the
mechanism giving rise to this operator, its associated mass scale or
its flavour structure. Its strength may be suppressed by a large scale
in the denominator (top-down) scenario, as in seesaw
schemes~\cite{Valle:2006vb}.  Alternatively, the strength may be
suppressed by small parameters (e.g. scales, Yukawa couplings) and/or
loop-factors (bottom-up scenario) with no need for a large scale, as
also reviewed in Ref.~\cite{Valle:2006vb}.

%--------------------------------------------------------------------------

\def\baselinestretch{1}
%\bibliographystyle{h-physrev4} 
%\bibliography{valle-ref,nu-rev06,parke-ref}

%\bibliography{lgenesis-ref}

\end{document}